# The Orbital Eccentricities of Binary Millisecond Pulsars in Globular Clusters

Frederic A. Rasio[1,2] and Douglas C. Heggie[3]

## ABSTRACT

Low-mass binary millisecond pulsars (LMBPs) are born with very small orbital eccentricities, typically of order $e_i \sim 10^{-6}$–$10^{-3}$. In globular clusters, however, higher eccentricities $e_f \gg e_i$ can be induced by dynamical interactions with passing stars. Here we show that the cross section for this process is much larger than previously estimated. This is because, even for initially circular binaries, the induced eccentricity $e_f$ for an encounter with pericenter separation $r_\mathrm{p}$ beyond a few times the binary semi-major axis $a$ declines only as a power-law, $e_f \propto (r_\mathrm{p}/a)^{-5/2}$, and *not* as an exponential. We find that *all* currently known LMBPs in clusters were probably affected by interactions, with their current eccentricities typically an order of magnitude or more greater than at birth.

*Subject headings:* binaries: close — celestial mechanics, stellar dynamics — globular clusters: general — pulsars: general

[1]Institute for Advanced Study, Olden Lane, Princeton, NJ 08540.

[2]Hubble Fellow.

[3]Department of Mathematics and Statistics, University of Edinburgh, King's Buildings, Edinburgh EH9 3JZ, Scotland.

## 1. Introduction

Low-mass binary millisecond pulsars (hereafter LMBPs) are thought to be formed when an old neutron star in a binary system accretes material from a red-giant companion. This leads to the production of a spun-up (recycled) pulsar with a low-mass white-dwarf companion in a wide, circular orbit. This formation process is well understood theoretically (for recent reviews, see Verbunt 1993; Phinney & Kulkarni 1994). The predicted relations between white-dwarf mass and orbital period (Joss, Rappaport, & Lewis 1987) and between orbital eccentricity and period (Phinney 1992) are in remarkable agreement with observations. The orbits tend to be very efficiently circularized by tidal torques and viscous dissipation in the red-giant envelope during the mass-transfer phase. Residual eccentricities $e_i \sim 10^{-6}$–$10^{-3}$ for LMBPs at birth are explained as arising from small fluctuations of the external gravitational field of the red giant caused by the convective motions in its envelope (Phinney 1992). In globular clusters, the wide orbits of LMBPs can be perturbed significantly by passing stars. Thus final eccentricities $e_f \gg e_i$ could be observed today, and the measured values contain important information about the dynamical history of the binaries and their environment. In dense clusters binaries can even be completely disrupted through interactions, so we expect to find LMBPs with eccentricities in the entire range $10^{-6} \lesssim e_f \lesssim 1$.

Using numerical scattering experiments, Rappaport, Putney, & Verbunt (1989) computed cross sections for inducing eccentricity in binaries with initially circular orbits. More recently, Phinney (1992) discussed the problem on the basis of approximate analytic expressions for the induced eccentricity derived from analytic perturbation theory. These two studies agree in predicting that the induced eccentricity decreases *exponentially* with increasing pericenter distance $r_p$. In contrast, Hut & Paczyński (1984), also on the basis of numerical work, found that the induced eccentricity decreased as a *power-law*, with $e_f \propto r_p^{-3}$, and speculated that this was coming from a non-vanishing higher-order secular perturbation effect (At lowest order secular perturbation theory gives a vanishing result for the induced eccentricity when the binary orbit is initially circular).

The purpose of this *Letter* is to shed new light on this question, by presenting and discussing our own analytic results for the induced eccentricity (§2). On the basis of detailed comparisons with numerical integrations, we believe these analytic results to be correct over the entire range of interest for LMBPs, $10^{-6} \lesssim e_f \lesssim 1$. The previously published expressions in which $e_f$ decreases exponentially with $r_p$ can grossly underestimate the effect when $e_f \lesssim 0.01$. Indeed, we find that over the intermediate range where $e_i < e_f \lesssim 10^{-2}$, the induced eccentricity generally varies as a power-law $e_f \propto r_p^{-5/2}$, making the cross section much larger than would be estimated on the basis of previous work. The implications of our results for LMBPs in globular clusters are discussed in §3.

## 2. Eccentricity Perturbations

For binaries with initial eccentricity $e_i \neq 0$, it is easy to show, in lowest order, that secular perturbation theory gives

$$\Delta e = e_f - e_i = -\frac{15\pi}{8\sqrt{2}} \frac{m_3}{M_{12}^{1/2} M_{123}^{1/2}} \left(\frac{a}{r_p}\right)^{3/2} e_i \sin(2\Omega)(1 - \cos^2 i). \qquad (1)$$

Here $m_1$ and $m_2$ are the masses of the binary components, $m_3$ is the mass of the passing star, $M_{12} = m_1 + m_2$, $M_{123} = M_{12} + m_3$, $a$ is the semi-major axis of the binary, $\Omega$ the longitude of the node (measured in the orbital plane of the binary from its pericenter), and $i$ is the inclination ($i = 0$ is coplanar prograde; $i = \pi$ is coplanar retrograde). Throughout this paper we assume for simplicity that $m_3$ is initially on a parabolic orbit with respect to the center of mass of the binary. A first immediate consequence of equation (1) is that, except for extreme mass ratios, one always has $|\Delta e|/e_i \ll 1$, as long as $r_p$ is at least a few times $a$. Thus, within the range of applicability of equation (1) (which includes the $r_p \to \infty$ limit; see below), encounters with passing stars produce a negligible change in eccentricity. Second, we see immediately that this lowest-order expression goes to zero for $e_i = 0$. We must therefore ask if higher-order terms could not become dominant over some intermediate range of values for $r_p$.

Indeed, a secular perturbation calculation at the next order, for $e_i = 0$, gives

$$e_f = \frac{15\pi\sqrt{2}}{64} \frac{m_3 |m_1 - m_2|}{M_{12}^{3/2} M_{123}^{1/2}} \left(\frac{a}{r_p}\right)^{5/2} \mathcal{D}(i,\omega), \qquad (2)$$

with

$$\mathcal{D}(i,\omega) = \left\{ \cos^2\omega \left[1 - \frac{5}{4}\sin^2 i\right]^2 + \sin^2\omega \cos^2 i \left[1 - \frac{15}{4}\sin^2 i\right]^2 \right\}^{1/2}. \qquad (3)$$

Here $\omega$ is the longitude of the pericenter of the third body, measured from the node. Notice that expression (2) vanishes in the equal-mass case ($m_1 = m_2$). A detailed derivation of expressions (2) and (3) will be given in Heggie & Rasio (1995; hereafter HR). The easiest derivation that we have found exploits the Lenz vector (or "eccentric axis"; cf. Pollard 1976). If only the quadrupole interaction is included the result is equation (1), whereas the octupole interaction gives equations (2)–(3) when $e_i = 0$.

Equations (1)–(3) were obtained from secular perturbation theory, i.e., by averaging over the orbital motion of the binary. This is a valid assumption for sufficiently large $r_p$, when the encounter is quasi-adiabatic. At smaller $r_\mathrm{p}$, however, the duration of the encounter can be comparable to the orbital period of the binary, and departures from adiabaticity can be important. Small deviations from adiabaticity can be calculated analytically using the method developed by Heggie (1975, §5.4) for estimating the changes in the semi-major axis of the binary (which vanish to all orders in secular perturbation theory). Using this method we obtain, for $e_i = 0$,

$$e_f = 3\sqrt{2\pi}\,\frac{m_3 M_{12}^{1/4}}{M_{123}^{5/4}}\left(\frac{2r_\mathrm{p}}{a}\right)^{3/4}\exp\left[-\frac{1}{3}\left(\frac{M_{12}}{M_{123}}\right)^{1/2}\left(\frac{2r_\mathrm{p}}{a}\right)^{3/2}\right]\mathcal{H}(i,\Omega,\phi),\qquad(4)$$

with

$$\mathcal{H}(i,\Omega,\phi) = \cos^2(i/2)\left\{\cos^4(i/2) + \frac{4}{9}\sin^4(i/2) + \frac{4}{3}\sin^2(i/2)\cos^2(i/2)\cos\left[2(\phi - 2\Omega)\right]\right\}^{1/2}.\qquad(5)$$

Here $\phi$ is the orbital phase of the binary at pericenter. Notice that this expression is maximum for a coplanar prograde orbit ($i = 0$), and vanishes for a coplanar retrograde orbit ($i = \pi$).

Our theoretical expressions (1)–(5) are in excellent quantitative agreement with the results of numerical integrations. Extensive comparisons will be presented in HR. Here, in Figure 1, we only illustrate this agreement for a typical LMBP case, and for $i = \Omega = \omega = 0$. The exponential expression (4) becomes important for $e_f \gtrsim 0.01$, but the power-law expression (2) dominates for $e_i \lesssim e_f \lesssim 0.01$. The dependence on phase is quite strong in the exponential regime, but becomes completely negligible in the power-law region. When combining expressions (1)–(5), we find that the best agreement with numerical results is obtained when we simply use the expression giving the largest $e_f$ for fixed $r_\mathrm{p}$.

## 3. Implications for Binary Pulsars

An immediate implication of our results, evident in Figure 1, is that cross sections for induced eccentricity can be over an order of magnitude larger than would be estimated on the basis of the exponential result (eq. [4]). This is for parabolic orbits, or equivalently, hard binaries (for which the cross section $\sigma \propto r_\mathrm{p}$). In general, for slightly hyperbolic orbits, the induced eccentricity at fixed $r_\mathrm{p}$ increases with relative velocity, and the relative increase in the cross section is even larger (Detailed analytic and numerical results for hyperbolic orbits will be presented in HR).

We can easily estimate the timescale $t_{>e}$ for inducing an eccentricity $e_f > e$. For $e_i < e \lesssim 0.01$, this is determined by setting the angle-averaged expression (2) equal to $e$. In the parabolic limit, and averaging over a Maxwellian distribution of relative velocities, we find

$$t_{>e} \simeq 4 \times 10^{11} \, n_4^{-1} \, v_{10} \, P_d^{-2/3} \, e^{2/5} \text{ yr} \qquad (e_i < e \lesssim 0.01) \qquad (6)$$

Here $n_4$ is the number density of stars (assumed for simplicity to be all of mass $m_3$) in units of $10^4 \, \text{pc}^{-3}$, $v_{10}$ is the one-dimensional velocity dispersion in units of $10 \, \text{km s}^{-1}$, $P_d$ is the LMBP's orbital period in days, and we have used the same typical values of the masses as in Figure 1 (the dependence on the masses in general is very weak over the relevant range for LMBPs in globular clusters). Deviations from the parabolic limit remain small for binaries with $P \lesssim 10^3$ days. For $e \gtrsim 0.01$ a similar procedure using expression (4) gives

$$t_{>e} \simeq 2 \times 10^{11} \, n_4^{-1} \, v_{10} \, P_d^{-2/3} \, [-\ln(e/4)]^{-2/3} \text{ yr} \qquad (e \gtrsim 0.01) \qquad (7)$$

Equations (6) and (7) can be used to estimate the age of a system with measured eccentricity. Equivalently, we can express the mean induced eccentricity in a system of age $t_9 \, 10^9$ yr as

$$e \simeq \text{Max} \left\{ \left(\frac{\eta}{400}\right)^{5/2} P_d^{5/3}, \; 4 \exp\left[-\left(\frac{\eta}{200}\right)^{-3/2} P_d^{-1}\right] \right\}, \qquad (8)$$

where we have defined $\eta = t_9 n_4 / v_{10}$. Since the cluster parameters $n$ and $v$ are usually known quite accurately, any uncertainty in $\eta$ comes primarily from the age of the system. The steep dependence of $e$ on $\eta$ suggests that eccentricity may be an excellent dynamical age estimator for LMBPs in globular clusters. This is important because the timing age of a cluster pulsar is usually unreliable, the measured pulse period derivative being contaminated by the acceleration of the pulsar in the mean gravitational potential of the cluster (Phinney 1992).

In Figure 2 we show the results of applying equation (8) to the nine LMBPs currently known in globular clusters. For all systems we calculate the range of expected values for the eccentricity corresponding to a pulsar age between $10^9$ yr and $10^{10}$ yr. All reliably measured timing ages of LMBPs (both in clusters and in the Galaxy) fall inside this range. Unless otherwise indicated below, all data are from the compilation by Phinney (1992, Tables 1 and 2). We also show for reference the contours of constant $\eta$ in the $(P, e)$ plane, calculated from equation (8), as well as the theoretically predicted "birth line," following the eccentricity-period relation for unperturbed systems (Phinney 1992). With very few exceptions Galactic LMBPs are observed to lie close to this line (Phinney & Kulkarni 1994).

For six out of nine currently known LMBPs in clusters, the orbital eccentricity has not yet been measured. For these systems, our results can be taken as crude order-of-magnitude predictions. We note that all these systems lie in the region of the plot where the use of the exponential result (eq. [4]) would lead to completely incorrect results (underestimating the predicted e by many orders of magnitude). Note also that all systems have a predicted eccentricity that is in principle measurable: values as small as $e \sim 10^{-6}$ are detectable by pulsar timing (PSR J2317+1439 has a measured eccentricity of $1.2 \times 10^{-6}$; Camilo, Nice, & Taylor 1993). In all cases where an upper limit on $e$ has been quoted by the observers, it is compatible with our predicted range. For the M53 binary, the current upper limit $e < 0.01$ (Kulkarni et al. 1991) falls just inside our predicted range and implies a maximum age of $9 \times 10^9$ yr for the pulsar.

Three systems have a measured and rather large eccentricity: PSR B1516+02B in M5 ($e = 0.13$; Wolszczan et al. 1989), PSR B1620-26 in M4 (e=0.025; Thorsett, Arzoumanian, & Taylor 1993), and PSR B1802-07 in NGC 6539 (e=0.21; D'Amico et al. 1993). The last two systems are thought to have more complicated dynamical histories than we have considered here. The M4 pulsar is actually in a hierarchical triple system which was probably formed through an exchange interaction involving the LMBP and another binary (Rasio, McMillan, & Hut 1995). The inner binary pulsar is likely to have its eccentricity perturbed mainly by the second companion in the current triple (Rasio 1994). For the NGC 6539 binary, the very large measured eccentricity, $e = 0.21$, is about two orders of magnitude above the maximum possible value induced by distant interactions with passing stars (obtained from eq. [8] when $t \simeq 10^{10}$ yr). In addition, this system also violates the predicted relation between companion mass and orbital period for a standard LMBP (Phinney 1992). Thus the present companion was probably acquired through a more violent interaction that left it in a highly eccentric orbit. The M5B binary also has a very large eccentricity, $e = 0.13$, but this value *is* consistent with having been induced by distant interactions with other stars in the cluster. Using equation (7) we predict an age $t = 7 \times 10^9$ yr and a pulse period derivative $\dot{P}_p < 2 \times 10^{-20}$ for the pulsar [assuming $t < P_p/(2\dot{P}_p)$].

We thank Sterl Phinney and Steve Thorsett for many useful discussions, and Steve McMillan for first pointing out to us the fundamentally different behavior of the induced eccentricity in the unequal-mass case. F. A. R. is supported by a Hubble Fellowship, funded by NASA through Grant HF-1037.01-92A from the Space Telescope Science Institute, which is operated by AURA, Inc., under contract NAS5-26555. D. C. H. thanks the Institute for Advanced Study, Princeton, for its hospitality while work on this project was being carried out.

– 7 –

Fig. 1.— Final orbital eccentricity $e_f$ following an interaction with a star on a coplanar prograde parabolic orbit with pericenter distance $r_p$ (given in units of the binary semi-major axis $a$). The masses of the two binary components are $m_1 = 1.4\,M_\odot$ and $m_2 = 0.2\,M_\odot$; the mass of the passing star is $m_3 = 1\,M_\odot$. Solid lines show the analytic results (eqs. [1]–[5]); dots show the results of numerical integrations. Round dots are for a system with initial eccentricity $e_i = 10^{-6}$, squares for $e_i = 10^{-5}$, and triangles for $e_i = 10^{-4}$. The "error bars" show the full extent of the dependence on the orbital phase. The dashed line shows the exponential decay predicted by eq. (4) alone.

Fig. 2.— Orbital period (in days) and eccentricity of the nine currently known LMBPs in globular clusters. The round dots show the positions of the three systems with measured values of both $e$ and $P$. The vertical dotted segments show the predicted eccentricity for a pulsar age between $10^9$ and $10^{10}$ years (eq. [8]). For the M53 binary we have taken into account the current upper limit ($e < 0.01$). Solid lines show the variation of $e$ with $P$ for fixed $\eta = t_9 n_4 / v_{10}$, and are logarithmically spaced with two lines per decade of $\eta$. The dashed curve shows the theoretical "birth line," along which unperturbed systems are expected to lie.

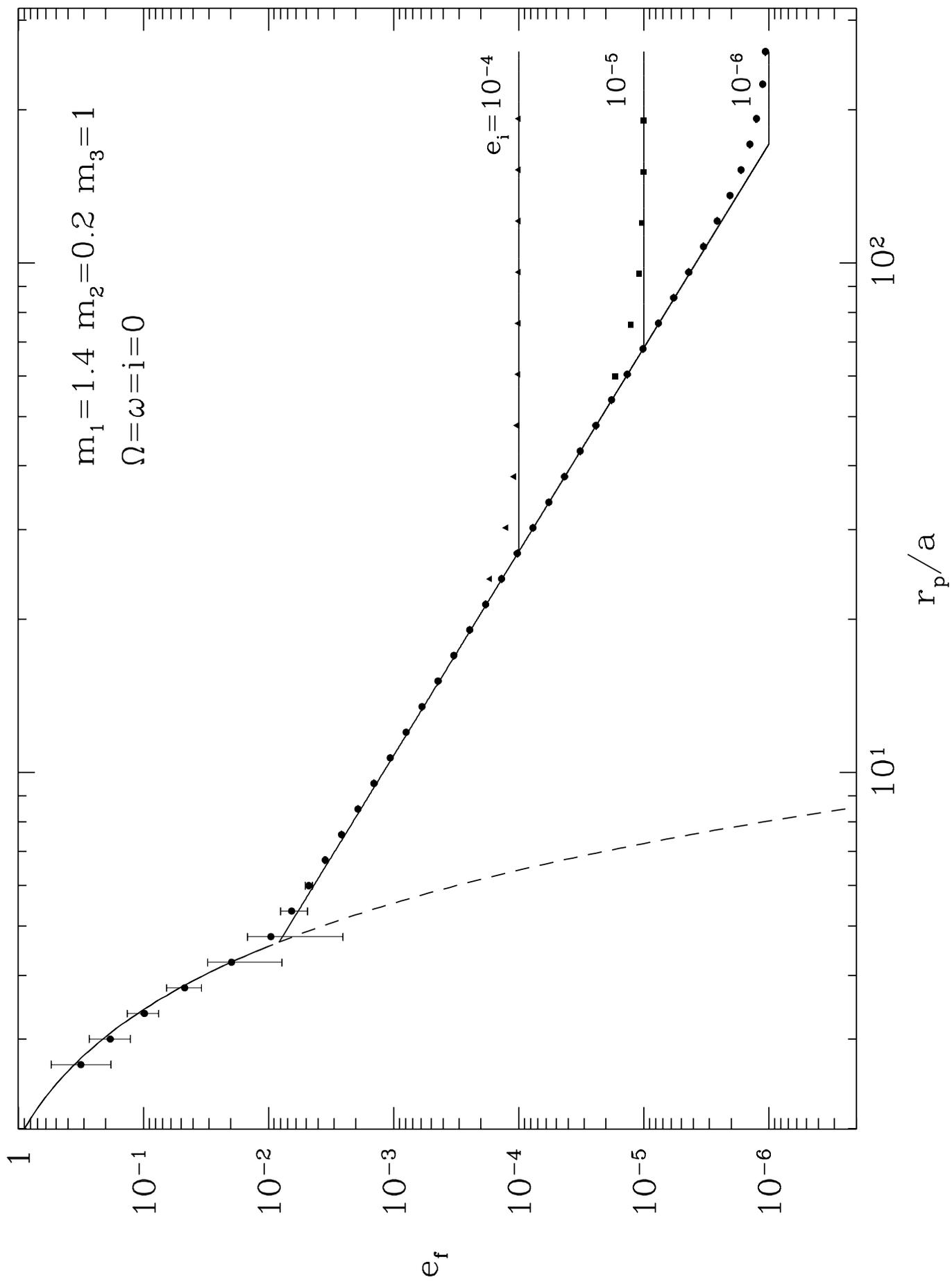

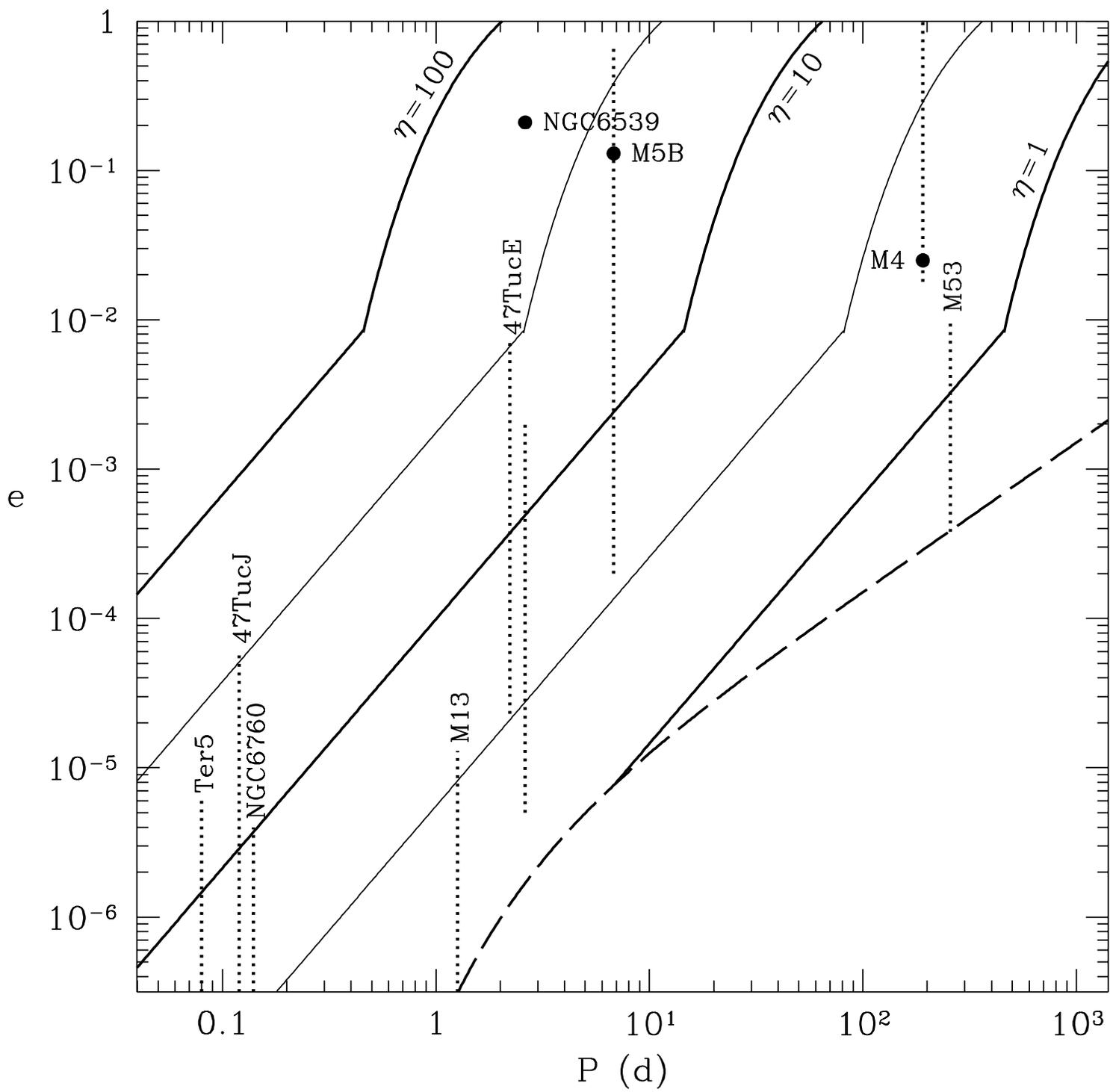